\documentclass[12pt]{article}
\usepackage{epsf}
\def\be{\begin{equation}}
\def\ee{\end{equation}}
\def\bea{\begin{eqnarray}}
\def\eea{\end{eqnarray}}

\begin{document}
\begin{titlepage}
\begin{center}
{\Large \bf William I. Fine Theoretical Physics Institute \\
University of Minnesota \\}
\end{center}
\vspace{0.2in}
\begin{flushright}
FTPI-MINN-15/09 \\
UMN-TH-3423/15 \\
March 2015 \\
\end{flushright}
\vspace{0.3in}
\begin{center}
{\Large  $X(3915)$ as a $D_s \bar D_s$ bound state.
\\}
\vspace{0.2in}
{\bf Xin Li$^a$  and M.B. Voloshin$^{a,b,c}$  \\ }
$^a$School of Physics and Astronomy, University of Minnesota, Minneapolis, MN 55455, USA \\
$^b$William I. Fine Theoretical Physics Institute, University of
Minnesota,\\ Minneapolis, MN 55455, USA \\
$^c$Institute of Theoretical and Experimental Physics, Moscow, 117218, Russia
\\[0.2in]

\end{center}

\vspace{0.2in}

\begin{abstract}
We suggest that the observed properties of the charmoniumlike resonance $X(3915)$ can possibly be explained if it is an $S-$wave molecular bound state of $D_s \bar D_s$ meson pair with binding energy about 18\,MeV. In particular, the decays of $X(3915)$ to pairs of non-strange $D$ mesons are suppressed and proceed at a rate comparable to that of the decay $X(3915) \to \omega J/\psi$, whose branching fraction can be as large as about 0.3. Other major types of decay of $X(3915)$ with a comparable (or slightly smaller) rate are the transition $X(3915) \to \eta \eta_c$ and the decays into light hadrons due to the annihilation of the $c \bar c$ quark pair. The existence of a  bound state should lead to an enhancement in the spectrum of the invariant mass for the $D_s \bar D_s$ near threshold in $B$ decays, e.g. in $B \to K D_s \bar D_s$ which enhancement can be tested experimentally.
\end{abstract}
\end{titlepage}

The charmoniumlike state $X(3915)$  presents a considerable challenge for understanding its internal structure. This resonance with the mass~\cite{pdg} $M=3918.4 \pm 1.9\,$MeV and width $\Gamma = 20 \pm 5\,$MeV has been observed only in the decay mode $X(3915) \to \omega J/\psi$ in $B$ decays, $B \to K \, X(3915) \,  \to K \, \omega J/\psi $~\cite{belleK,babarK}, with the combined branching fraction measured as
\be
{\cal B} \left[ B^+ \to K^+ \, X(3915) \right ] \, {\cal B} \left[ X(3915) \to \omega J/\psi \right ] = \left ( 3.0^{+0.9}_{-0.7} \right ) \times 10^{-5}~,
\label{bkx}
\ee
and in two photon production~\cite{belleg,babarg} with the yield described by~\cite{pdg} 
\be
\Gamma \left [ X(3915) \to \gamma \gamma \right ] \, {\cal B} \left[ X(3915) \to \omega J/\psi \right ] = 54 \pm 9 \, {\rm eV}~.
\label{gx}
\ee
Furthermore, a BaBar angular analysis~\cite{babarg} of the two photon production favors the spin - parity assignment $J^P=0^+$ for $X(3915)$. For this reason an assignment~\cite{lls} of this resonance as a radially excited charmonium $^3P_0$ state $\chi_{c0}(2P)$ was hastily adopted by BaBar~\cite{babarg} and later by PDG~\cite{pdg}. However it has been convincingly  argued~\cite{gm,Olsen} that the observed properties of the resonance $X(3915)$ are highly unlikely for a $\chi_{c0}(2P)$ state, suggesting~\cite{gm} that the actual $\chi_{c0}(2P)$ lies at the mass of about 3840\,MeV and has a broad width, $\Gamma \sim 200\,$MeV dominated by the decay to $D \bar D$ pairs, and thus leaving the resonance $X(3915)$ to be interpreted as an exotic state. In this paper we discuss the possibility that this resonance is a bound state of a $D_s \bar D_s$ meson pair, whose threshold is the nearest in mass to $X(3915)$. We argue that such interpretation might explain, although currently with a considerable uncertainty, the unusual observed properties of this resonance  and that it can be tested experimentally, e.g. by studying the decays $B \to K\, D_s \bar D_s$ near the threshold of the strange charmed meson-antimeson pair.  In what follows we briefly review the peculiarities in the measured properties of $X(3915)$ pointed out in Refs.~\cite{gm,Olsen} that, in particular, preclude the $\chi_{c0}(2P)$ assignment and then discuss how those properties can be implemented in the $D_s \bar D_s$ molecular model.

A major peculiarity of the resonance $X(3915)$ is the large combined branching fraction in Eq.(\ref{bkx}) considered together with the general understanding that for a pure charmonium the rates of transitions with emission of light hadrons are typically quite small with the largest rate [that of $\psi(2S) \to \pi \pi \, J/\psi$ ] being of the order of 0.1\,MeV. Indeed, if one conservatively caps the rate of the decay $X(3915) \to \omega J/\psi$ at 1\,MeV, corresponding to ${\cal B} \left[ X(3915) \to \omega J/\psi \right ] < 5\% $, the branching fraction for the $B$ decay, ${\cal B} \left[ B^+ \to K^+ \, X(3915) \right ]$ would be estimated as being larger than approximately $6 \times 10^{-4}$, which would make the $X(3915)$ one of the most abundantly produced charmonium states in the decays of $B$ meson. In particular, this estimate would exceed the measured yield of the $\chi_{c0}(1P)$ state,  ${\cal B} \left[ B^+ \to K^+ \, \chi_{c0}(1P) \right ] = (1.50^{+0.15}_{-0.14}) \times 10^{-4}$, which is hardly acceptable in any reasonable theoretical scheme. 

Another unusual property of $X(3915)$ is the apparent absence of the dominance of its decay into $D \bar D$ meson pairs. Namely, the current measurements show no signal for such decay, and using the available Belle data~\cite{belledd} Olsen~\cite{Olsen} estimates ${\cal B} \left [  X(3915) \to D^0 \bar D^0 \right ] < 1.2 \, {\cal B} \left[ X(3915) \to \omega J/\psi \right ]$, which, given the isotopic symmetry, translates to
\be
{\cal B} \left [  X(3915) \to D \bar D \right ] < 2.4 \, {\cal B} \left[ X(3915) \to \omega J/\psi \right ]~.
\label{bd}
\ee
This behavior is in a remarkable contrast with expectation for a $0^+$ charmonium and with the one observed for the spin-2 resonance identified~\cite{pdg} as $\chi_{c2}(2P)$ with mass $3927.2 \pm 2.6\,$MeV, whose width $\Gamma = 24 \pm 6\,$MeV is practically saturated by its decay into $D-$wave $D \bar D$ meson pairs.~\footnote{The very small mass splitting of only about 9\,MeV between the $\chi_{c2}$ and $X(3915)$ is used~\cite{gm} as an additional argument against the interpretation of $X(3915)$ as $\chi_{c0}(2P)$ charmonium state.}

Clearly, the conclusion regarding an unacceptably high rate of the decay $B \to K \, X(3915)$ derived from the experimental number in Eq.(\ref{bkx}) is invalidated if the branching fraction  ${\cal B} \left[ X(3915) \to \omega J/\psi \right ]$ is significantly larger than 5\%, say about 30\%, while the upper bound in Eq.(\ref{bd}) could be satisfied if one could identify a mechanism for suppression of the decay $X(3915) \to D \bar D$. We will argue here that both these requirements might be fulfilled if the resonance $X(3915)$ is an $S-$wave bound molecular state of the $D_s \bar D_s$ meson pair. The binding energy $\Delta$ in this case is 
\be
\Delta = 2 M(D_s) - M_X = 18.2 \pm 1.9\, {\rm MeV}~,
\label{de}
\ee
so that the bound state is moderately shallow. 
 
The argument involving the perceived relative suppression of the decay $X(3915) \to \omega J/\psi$ is based on invoking the Okubo-Zweig-Iizuka (OZI) rule. For a $D_s \bar D_s$ state both this decay and the decay to $D \bar D$ would be OZI suppressed and may thus have a comparable rate. As for the absolute rate of $X(3915) \to \omega J/\psi$, it can be recalled that the original application of the OZI rule was for an explanation of the suppression of the decay $\phi \to \rho \pi$. While the latter decay is indeed OZI suppressed relative to the decay into $K \bar K$, in absolute terms its width is approximately 0.65\,MeV, even though this is a $P-$wave decay with a moderate c.m. momentum of 179\,MeV. The decay $X(3915) \to \omega J/\psi$ is an $S-$wave process where, in the discussed picture, the $s \bar s$ pair converts to $\omega$  with no apparent kinematical suppression since the momentum in this decay is about 220\,MeV. It thus does not seem unreasonable to suggest that even though the latter decay is OZI suppressed, similarly to $\phi \to \rho \pi$, the $S-$wave amplitude is enhanced in comparison with the $P-$wave in the $\phi$ meson decay, so that an order of magnitude larger decay rate, $\Gamma \left[ X(3915) \to \omega J/\psi \right ] \sim 6\,$MeV, required for making the branching fraction of approximately 30\%, might in fact be not untypical. 

The decay into the pairs of non-strange $D$ mesons, $X(3915) \to D \bar D$, is similarly OZI suppressed, so that it should come as no surprise that its rate is comparable to that of the decay to $\omega J/\psi$. Furthermore the OZI rule also keeps $X(3915)$ as a $D_s \bar D_s$ molecule from a significant mixing with $^3P_0$ states of charmonium, which mixing would also certainly give rise to enhanced decays into $D \bar D$.

A possible caveat in the discussed scheme for $X(3915)$ and its decays is related to the decay $X(3915) \to \eta \eta_c$, which in fact is the only possible process for a $D_s \bar D_s$ molecule that is not OZI suppressed. A recent search for this process by Belle~\cite{belleeta} resulted in the upper bound (at 90\% CL)
\be
{\cal B} \left[ B^+ \to K^+ \, X(3915) \right ] \, {\cal B} \left[ X(3915) \to \eta \eta_c \right ] < 3.3 \times 10^{-5}~.
\label{bkxe}
\ee
By comparing this bound with the number in Eq.(\ref{bkx}) one can readily see that the decays $X(3915) \to \omega J/\psi$ and $X(3915) \to \eta \eta_c$ can have comparable rates. It should be pointed out that there is at least one possible factor that somewhat mitigates the OZI suppression of the former process in comparison with the latter. This factor is related to the heavy quark spin structure of an $S-$wave of two pseudoscalar mesons $D_s \bar D_s$. In the mesons the spin of the heavy quark (antiquark) is fully correlated with the spin of the light antiquark (quark). As a result the meson-antimeson system is a mixed state with regards to the total spin of the $c \bar c$ pair. The decomposition in terms of the states with definite spin of the heavy pair, $S_H$ and the light quark-antiquark pair $S_L$ is well known~\cite{mv}. For the $S-$wave pair of pseudoscalar mesons it reads as
\be
\left ( D_s \bar D_s \right )_{0^+} = {1 \over 2} \, 0_H \otimes 0_L - {\sqrt{3} \over 2} \, 1_H \otimes 1_L~,
\label{shl}
\ee
so that in the discussed $D_s \bar D_s$ molecule the probability for the $c \bar c$ pair to be in a spin triplet state is 3/4, while the probability for a spin singlet is 1/4. Since the heavy quark spin is approximately conserved, the transition to the spin triplet $J/\psi$ level is enhanced by a factor of three over that to the spin singlet $\eta_c$. Furthermore, the OZI allowed transition of the $s \bar s$ pair to the $\eta$ meson, $\eta = (  u \bar u + d \bar d - 2 s \bar s)/\sqrt{6}$ in the limit of flavor $SU(3)$ symmetry, carries another factor of 2/3 in the probability. As a result the ratio of the transition rates can be factorized as
\be
{\Gamma(D_s \bar D_s \to \eta \eta_c) \over \Gamma(D_s \bar D_s \to \omega J/\psi)} = {2 \over 9} \, {p_\eta \over p_\omega} \, {1 \over R} \, \left | {F(p_\eta) \over F(p_\omega)} \right |^2~,
\label{eor}
\ee
where $R$ is the OZI suppression factor, $p_\omega \approx 220\,$MeV ($p_\eta \approx 665\,$MeV) is the momentum of $\omega$ ($\eta$) in the decay, and $F(p)$ is the form factor accounting for the absorbed recoil momentum by the charmonium ground state. One can see that, although a larger momentum in the $\eta$ transition gives a larger phase space, the reduction of the amplitude by the form factor $F(p)$ can be more significant, since $p_\eta$ is comparable with the typical momentum in the ground state charmonium. Clearly, lacking a better description of the discussed transitions, it appears impossible to reliably estimate the ratio of the rates in Eq.(\ref{eor}), and the question as to whether the $\eta$ transition can have a rate not larger than the decay to $\omega J/\psi$ remains open, and one has to turn to other ways of testing the suggested hidden strangeness molecular structure of $X(3915)$.

One characteristic signature of a bound $D_s \bar D_s$ $S-$wave state arises if one treats it as shallow bound state dominating the near threshold  $D_s \bar D_s$ scattering amplitude. At the binding energy as in Eq.(\ref{de}) the characteristic under-threshold momentum is $\kappa = \sqrt{M_D \Delta} \approx 190\,$MeV, so that the amplitude can be described by the scattering length
\be
a = {1 \over \kappa} \approx 1\,{\rm fm}~,
\label{sla}
\ee
which also determines the characteristic distances at which the mesons move in the bound state. It should be mentioned that, most likely, the 1 fermi length scale is near the limit at which the effective radius approximation can be applied to the amplitude, or, in other words,  the bound state at 18\,MeV is at the limit of being considered as `shallow'. Also such distances are comparable with the size of the mesons, so that considering them as point objects is only marginally applicable. However in lieu of a more detailed approach we use here the approximation of a `large' scattering length for description of the near threshold dynamics of a $D_s \bar D_s$ pair. In this approximation the $D_s \bar D_s$ scattering amplitude at a momentum $k$ above the threshold can be written as
\be
f=- {1 \over \kappa + i \gamma + i k}
\label{amp}
\ee
where the imaginary shift $i \gamma$ for the position of the pole in $k$ takes into account the inelasticity due to the width $\Gamma$ of the resonance $X(3915)$, $\kappa+i \gamma = \sqrt{M_D (\Delta + i \Gamma/2)}$ (numerically $\gamma \approx 50\,$MeV). 

In the processes, where the $D_s \bar D_s$ pair is produced by a source localized at distances shorter than $a$, the production near the threshold can be considered as dominated by a resonance process, and the usage of the resonance amplitude in the form (\ref{amp}) is sometimes referred to as the Fatt\`e parametrization~\cite{flatte}. Applying this parametrization to the decays $B \to K D_s \bar D_s$, one readily finds that the distribution in the invariant mass $M_{D \bar D}$ of the $D_s \bar D_s$ pairs near the threshold is given by
\be
{d \over d M_{D \bar D}} \Gamma(B \to K D_s \bar D_s) = C \, { k \over \kappa^2 + (\gamma + k)^2} 
\label{flamp}
\ee
where $C$ is a constant and $k = \sqrt{M_D (M_{D \bar D} - 2 M_D)}$ is the relative momentum of the mesons in their center of mass frame. In the expression (\ref{flamp}) we neglect the effect of the Coulomb interaction between the mesons, which effect itself depends on the meson-antimeson strong interaction~\cite{lv} and becomes significant only in the immediate vicinity of the threshold. Clearly, the resonance corresponding to a shallow bound state enhances the yield of the meson pairs near the threshold as illustrated in Fig.1 and such enhancement can be tested in the existing or future data on the $B$ decays.

\begin{figure}[ht]
\begin{center}
 \leavevmode
    \epsfxsize=11cm
    \epsfbox{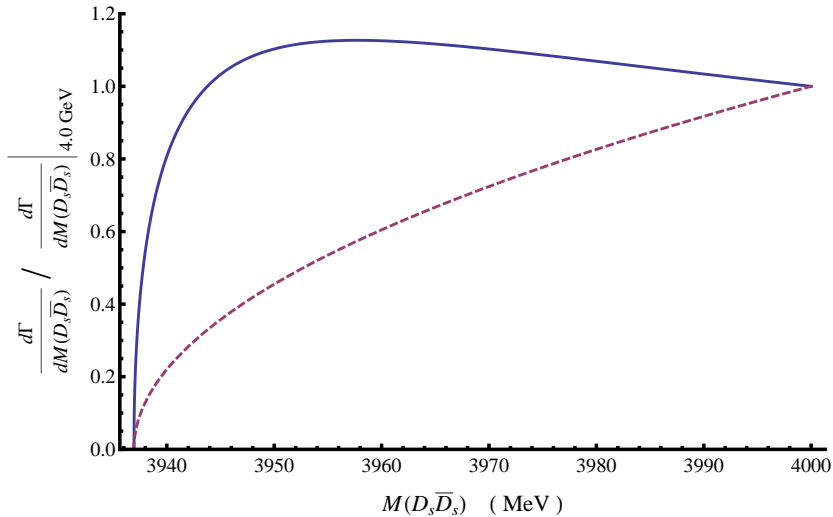}
    \caption{The spectrum of invariant mass for the $D_s \bar D_s$ system near the threshold with the enhancement due to a bound state (solid) as described by Eq.(\ref{flamp}), compared with the $S-$wave phase space (dashed). Both curves are normalized to the same value at $M(D_s \bar D_s)=4.0\,$GeV. }
\end{center}
\end{figure} 

According to the discussed picture of the $X(3915)$ resonance, the bulk of its total width is thus the sum of the decay rates to $\omega J/\psi$, $\eta \eta_c$ and $D \bar D$ all contributing a comparable amount. One more type of decay, possibly contributing a  slightly smaller but still noticeable amount to the total width is the decay to light hadrons due to the annihilation of the charmed quark and antiquark. In order to estimate (very approximately) the rate of such decays we start with a discussion of $X(3915)$ decay into two photons whose rate enters Eq.(\ref{gx}). For an $S-$wave nonrelativistic $D_s \bar D_s$ meson pair in a state described by a wave function $\psi(r)$ the width for annihilation into $\gamma \gamma$ is given by
\be
\Gamma (D_s \bar D_s \to \gamma \gamma) = {2 \pi \, \alpha^2 \over M_D^2} \, |\Phi|^2 \, |\psi(0)|^2~,
\label{2g}
\ee
where $\Phi$ is the form factor for the vertex $D_s \bar D_s \gamma \gamma$, normalized so that $\Phi=1$ for point particles of unit charge. Using for an (very approximate) estimate $|\psi(0)|^2 \sim \kappa^3/\pi$ in the bound state $X(3915)$, one gets
\be
\Gamma[X(3915) \to \gamma \gamma] \sim 0.2 \, |\Phi|^2 \,{\rm keV}
\label{2gx}
\ee
which is in the same ballpark as the experimental number (\ref{gx}), if $|\Phi|$ is not significantly less than one and if, as we discuss here, ${\cal B} \left[ X(3915) \to \omega J/\psi \right ] \sim 0.3$. 

The annihilation of the $c \bar c$ pair from the molecular state can proceed either through two gluons for the heavy quark pair in a spin singlet state, or through a one-gluon mediated process $c \bar c \to q \bar q$ for the $c \bar c$ pair in a color-octet spin-triplet state. For the known charmonium states: $\eta_c$, $\chi_{c0}$, $\chi_{c2}$, that decay both to two photons and to light hadrons through two gluons, the ratio of the $\gamma \gamma$ decay rate to that of the hadronic decay is around $2 \times 10^{-4}$. Allowing for the additional contribution of the annihilation through one gluon from the molecular state, it appears reasonable to approximately estimate the similar ratio for $X(3915)$ as
\be
{\Gamma[X(3915) \to \gamma \gamma] \over \Gamma[X(3915) \to {\rm light ~ hadrons}]} \sim 10^{-4}~,
\label{xgg}
\ee
so that in absolute terms the annihilation width of $X(3915)$ is likely to be about (1.5 - 2)\,MeV and contribute about 10\%, or slightly less to its total width. Naturally, one should expect, due to the presence of the $s \bar s$ quark pair in $X(3915)$,  an enhanced $K/\pi$ ratio in such annihilation decays. However it is not clear whether this enhancement can be detected over the background, since these decays contribute only a rather small fraction of the total width of the resonance.

It should be also mentioned that an attraction between the strange charmed  $D_s$ and $\bar D_s$ mesons, resulting in the suggested bound state very likely implies an existence of even stronger binding for non-strange $D$ and $\bar D$ mesons in the isoscalar channel. (The isovector states of such type are discussed for the bottomonium sector in Ref.~\cite{mv}.) However the experimental signature and the very existence of an isoscalar $0^+$ bound state of $D \bar D$ is most probably strongly affected by mixing with $^3P_0$ charmonium, and a discussion of such state is certainly beyond our present scope. 

In summary. We argue that the unusual observed properties of the $X(3915)$ resonance, in particular its apparently large decay rate to $\omega J/\psi$ and a suppressed decay into $D \bar D$ pairs may possibly be understood if this resonance is a molecular bound state of a $D_s \bar D_s$ meson pair. Unavoidably, our estimates of the properties of such molecular object are subject to large uncertainties because of the current lack of a reliable description of the strong multiquark dynamics. However the possibility that $X(3915)$ is an under threshold pole in the $D_s \bar D_s$ scattering amplitude located close to the physical region can be tested experimentally by searching for the enhanced yield of these meson-antimeson pairs near the threshold, e.g. in the decays $B \to K D_s \bar D_s$. 

M.~B.~V. thanks Roman Mizuk for enlightening discussions of the experimental data related to the $X(3915)$ resonance. The work of M.~B.~V.  is supported, in part, by U.S. Department of Energy Grant No. DE-SC0011842.

\end{document}